\documentclass[apj,twocolumn,breaklinks,colorlinks,citecolor=blue,urlcolor=blue]{openjournal}

\usepackage{amsmath}
\usepackage{amssymb}
\usepackage[rgb]{xcolor}
\definecolor{MyGreen}{rgb}{0.0,0.6,0.3}
\definecolor{MyPurple}{rgb}{0.6,0,0.3}
\usepackage{csquotes}
\usepackage{color,colortbl}
\usepackage{orcidlink}

\begin{document}
\title{Early Post Asymptotic Giant Branch Instability: Does it Affect White Dwarf Hydrogen Envelope Mass?}
\author{James MacDonald}
\email{jimmacd@udel.edu}
\affiliation{Department of Physics and Astronomy, University of Delaware, Newark, DE 19716, USA}

\begin{abstract}
 Although most white dwarf stars have hydrogen-dominated atmospheres, a significant fraction have atmospheres in which hydrogen is spectroscopically absent, with the fraction of hydrogen-free atmospheres varying with effective temperature. Estimates of the total mass of hydrogen, $M_H$, in the stellar envelope from either asteroseismology or spectral evolution are at odds with predicted values from theoretical stellar evolution modeling. Recent work has found that models in the early post Asymptotic Giant Branch (AGB) phase of evolution can exhibit thermally and dynamical unstable behavior. Here we investigate whether this Early Post AGB Instability (EPAGBI) can help resolve the conflict in $M_H$ values determined from white dwarf spectral evolution, analysis of DAV pulsations and canonical stellar evolution modeling, by evolving solar composition models of mass $1$ and $2 \> \mathrm{M}_\odot$ through the AGB phases and to the white dwarf cooling track. We present results for when the EPAGBI phase of evolution is followed in detail, and for comparison purposes, when it is suppressed by forcing the time step to be large compared to the growth time of the instability. The $M_H$ values at the end of the calculations are in the range consistent with asteroseismological determinations. However, we caution that, because hydrodynamic behavior is not included in our modeling, it is possible that all hydrogen would be removed. Thus, it is unclear whether the occurrence of the EPAGBI resolves the discrepancy between predictions of stellar evolution modeling and the asteroseismological hydrogen envelope mass determinations. The major impact of EPAGBIs is that they cause loops in the HRD, which are absent when the EPAGBI is suppressed. For models of AGB-departure mass 0.567 and 0.642 $\mathrm{M}_\odot$, it takes approximately 100 and 10 yr for a single HRD loop, respectively. Such loops might be detectable in a long-term monitoring program, or perhaps by their imprint on planetary nebula morphology imparted by the cyclically varying mass loss rate. Since the characteristic timescale of the looping in the HRD depends on the stellar mass, if measurable, it could provide a way to determine the stellar mass just after AGB departure, particularly if it is $\gtrsim 0.72 \> \mathrm{M}_\odot$ for which we estimate a 1 yr loop timescale. Another signature of the EPAGBI is the production of lithium by the Cameron-Fowler process. During the EPAGBI phase the photospheric temperature is always much higher than the temperatures of stars of appropriate log \textit{g} for which the Li I resonance line can be detected, and Li detection is unlikely to be a way to identify the EPAGBI phase. However, the Li precursor, $^7$Be, is convected to the photosphere in significant amounts (up to  $\sim 400$ times the solar photospheric mass fraction) at various times in the EPAGBI phase, which may be detectable by observing the Be II  resonance doublet.

\end{abstract}


\maketitle

\section{Introduction}
\label{sec:intro}

Spectroscopic measurements have revealed that although most white dwarf stars (WDs) have hydrogen-dominated atmospheres, a significant fraction have atmospheres in which hydrogen is spectroscopically absent. Furthermore the ratio is found to vary with effective temperature, $T_\mathrm{eff}$. The spectroscopic evolution of WDs has recently been reviewed by \cite{2024Ap&SS.369...43B}. Although DA white dwarfs with $T_\mathrm{eff} \gtrsim 90,000$ K have been found, many hydrogen-atmosphere WDs contain traces of helium and hence are classified as type DAO. Also, for synthetic spectra to fit the Balmer line profiles, the presence of CNO elements seems required \citep{2010ApJ...720..581G}. A likely mechanism for the presence of helium is a stellar wind driven by radiation acting on heavy element ions \citep{2000A&A...359.1042U}. Due to the incompletely understood physically processes acting at the high temperatures, \cite{2024Ap&SS.369...43B} conclude that  determining the fraction of helium-atmosphere WDs, $f_{He}$ is unreliable for $T_\mathrm{eff} \gtrsim 90,000$ K. Initially, it was found that all WDs with $45,000$ K $\gtrsim T_\mathrm{eff} \gtrsim 30,000$ K had hydrogen-dominated atmospheres leading to the \enquote{DB Gap} \citep{1985ApJS...58..379W, 1986ApJ...309..241L}. Between 80,000 K and 45,000 K, the ratio of helium-rich to hydrogen-rich WDs decreases monotonically with decreasing $T_\mathrm{eff}$. More recently \cite{2020ApJ...901...93B} have shown that the fraction of helium-atmosphere WDs gradually decreases from $ \sim 24\%$ at $T_\mathrm{eff} \gtrsim 70,000$ K to $\sim 8\%$ at $T_\mathrm{eff} \sim 30,000$ K.

Recent studies have shown that the fraction of helium-atmosphere WDs then remains roughly constant down to $T_\mathrm{eff} \sim 20,000$ K and increases again at lower temperatures, with the incidence of helium-dominated atmospheres reaching $\sim 30\% $ at the end of the cooling sequence \citep{2022A&A...658A..79L}.

From these observations, \cite{2023ApJ...946...24B} conclude that (1) $\sim 15\% -30\%$ of WDs initially have a helium-rich atmosphere but experience a helium-to-hydrogen transition at high temperature and a hydrogen-to-helium transition at low temperature, (2) $\sim 5\%-10\%$ of WDs retain a helium-rich surface throughout their entire evolution, and (3) the remaining objects likely always possess a standard hydrogen-dominated atmosphere. 

To explain the evolution of the helium atmosphere ratio, \cite{1987fbs..conf..319F} proposed the 'float-up' model in which a small amount of hydrogen is initially distributed in the helium-rich envelope. As the star cools, hydrogen floats to the surface due to the effects of gravitational settling and element diffusion, and eventually forms a thin layer of pure hydrogen. Subsequent evolution depends on the mass of hydrogen, $M_H$ \citep{1991ApJ...371..719M, 2018ApJ...857...56R}. If $M_H \lesssim 10^{-14} \> \mathrm{M}_\odot$ the underlying helium layer becomes convectively unstable as the star cools, so that the thin hydrogen layer is eventually mixed into the much more massive helium layer. However, if $M_H \gtrsim 10^{-14} \> \mathrm{M}_\odot$, it is the hydrogen layer that becomes convectively unstable. As the star cools, the base of the convective layer moves deeper into the star and if $M_H \lesssim 10^{-8} - 10^{-7} \> \mathrm{M}_\odot$ reaches the helium-rich layer. At which point the photosphere becomes helium dominated. The surface temperature at which the transition occurs is strongly dependent on the value of $M_H$.

The idea that most WDs start cooling with helium dominated atmospheres in which a small amount of hydrogen is distributed is at odds with the findings of stellar evolution studies which indicate that the majority of WDs should begin cooling with a hydrogen-rich envelope of mass of ${\sim}10^{-4} \> \mathrm{M}_\odot$ \citep{1981A&A...103..119S, 1984ApJ...282..615I, 1985ApJ...296..540I}. Approximately 20 - 30 $\%$ of the Central Stars of Planetary Nebulae (CSPNe) are found to be hydrogen deficient \citep{2020A&A...640A..10W}. One mechanism for the formation of hydrogen-deficient CSPNe is the 'Born-Again' scenario \citep{1983ApJ...264..605I, 1984ApJ...277..333I} in which a Very Late Thermal Pulse (VLTP) occurs \citep{1977PASJ...29..331F, 1979A&A....79..108S, 1984ApJ...277..333I, 1995LNP...443...48I}. During the VLTP, hydrogen is ingested by the helium-flash-driven convection zone, leading to a Double Shell Flash (DSF) \citep{2008A&A...490..769C}. In principle, due to the large hydrogen-burning luminosity experienced in the DSF, the hydrogen-rich envelope could be completely ejected or sufficiently reduced in mass that radiatively-driven winds could remove any remaining hydrogen containing layer. While \cite{2024arXiv241118035M} have noted that many stellar models have found it difficult to remove all hydrogen from the star, leaving at least $10^{-11} \> \mathrm{M}_\odot$ remaining for a final WD mass of $\sim 0.6 \> \mathrm{M}_\odot$ \citep{2017ASPC..509..435M}, \cite{2006MNRAS.371..263L} find that in models where a helium shell flash occurs after the end of the Asymptotic Giant Branch (AGB) phase it is possible to remove essentially all hydrogen except for a small amount of ${\sim}10^{-16} \> \mathrm{M}_\odot$.

Asteroseismic studies of DAV (ZZ Ceti) stars place strong constraints on the hydrogen-rich envelope mass of relatively cool WDs ($T_\mathrm{eff} \simeq 12,000$ K. 
\cite{2009MNRAS.396.1709C} performed asteroseismic analysis of 83 DAV stars and found a range of values of log $(M_H/M_*)$ from  $-9.5$ to $-4$ with a mean of $-6.3$.
\cite{2012MNRAS.420.1462R} performed asteroseismic analysis of 44 ZZ Ceti stars. They concluded (1) 'There exists a range of thicknesses of the H envelope in the studied ZZ Ceti stars, in qualitative agreement with the results of \cite{2009MNRAS.396.1709C}. Our distribution of H envelope thicknesses is characterized by a strong peak at thick envelopes [$\log(M_H/M_*) \sim -4.5$] and another much less pronounced peak at very thin envelopes [$\log(M_H/M_*) \sim -9.5$], with an evident paucity for intermediate thicknesses.' and (2) 'In most of the analysed DAVs (34 stars from a total of 44), our asteroseismological models have H envelopes thinner than the values predicted by standard evolutionary computations for a given stellar mass'. More recently, \cite{2023ApJ...948...74H} have performed an asteroseismic analysis of 29 DAVs in the Kepler and K2 data sets. They also find a range in values of $\log(M_H/M_*)$ with 4 stars having values of $\log(M_H/M_*) \sim -4.5$ consistent with the thick envelopes predicted by standard evolution, but the others have thinner envelopes with $\log(M_H/M_*)$ as low as -9.25. Since there is a well-defined ZZ Ceti instability strip in the $T_\mathrm{eff}$ - log \textit{g} plane which contains all DA pulsators and no DA non-pulsators \citep{2004ApJ...600..404B}, the discrepancy between $M_H$ values predicted by stellar evolution calculations and those measured asteroseismicly cannot be explained by some of the thick hydrogen envelope DA being non-pulsators.

By hydrostatic and dynamical stellar evolution modeling, \cite{2023arXiv230311374G} has studied the development of rapidly growing radial pulsations near the end and after the thermally pulsing AGB phase. In certain cases, an instability develops after the model leaves the AGB and reaches $T_\mathrm{eff} \sim 4500$ K. The pulsation driving is found to lie in the partial ionization zones of hydrogen and helium. \cite{2023arXiv230311374G} also showed that the pulsation behavior could be accurately followed up to a point by quasi-hydrostatic equilibrium calculations, and as such are not excited acoustic waves, in contrast to classical pulsators. The periods of the pulsations were found to be in the range of about 100 days to over 1000 days, with e-folding times of a few periods' length which is much shorter than in classical pulsators.

In this paper, we examine whether this Early Post AGB Instability (EPAGBI) can help resolve the conflict in $M_H$ values determined from white dwarf spectral evolution, analysis of DAV pulsations and canonical stellar evolution modeling.

\section{Post AGB Stellar evolution models}
\label{sec:models}

We use the DEUCES code described by \cite{2023MNRAS.525.4700L} to evolve models of initial mass $1 \> \mathrm{M}_\odot$ and $2 \> \mathrm{M}_\odot$ from the pre-main sequence phase to the start of the WD cooling track. Models are computed with and without inclusion of  convective overshooting using the prescription of \cite{1997MNRAS.285..696S}. The main reason for inclusion of convective overshooting is to increase the chance of there being an EPAGBI phase as other instabilities can occur during the thermal pulse cycle [see \cite{1994A&A...290..807W}] that prevent continuing evolution past the end of the AGB.

\subsection{Evolution of the \texorpdfstring{$1 \> \mathrm{M}_\odot$}{} model}

\begin{figure}
\centering
\includegraphics[scale=0.3]{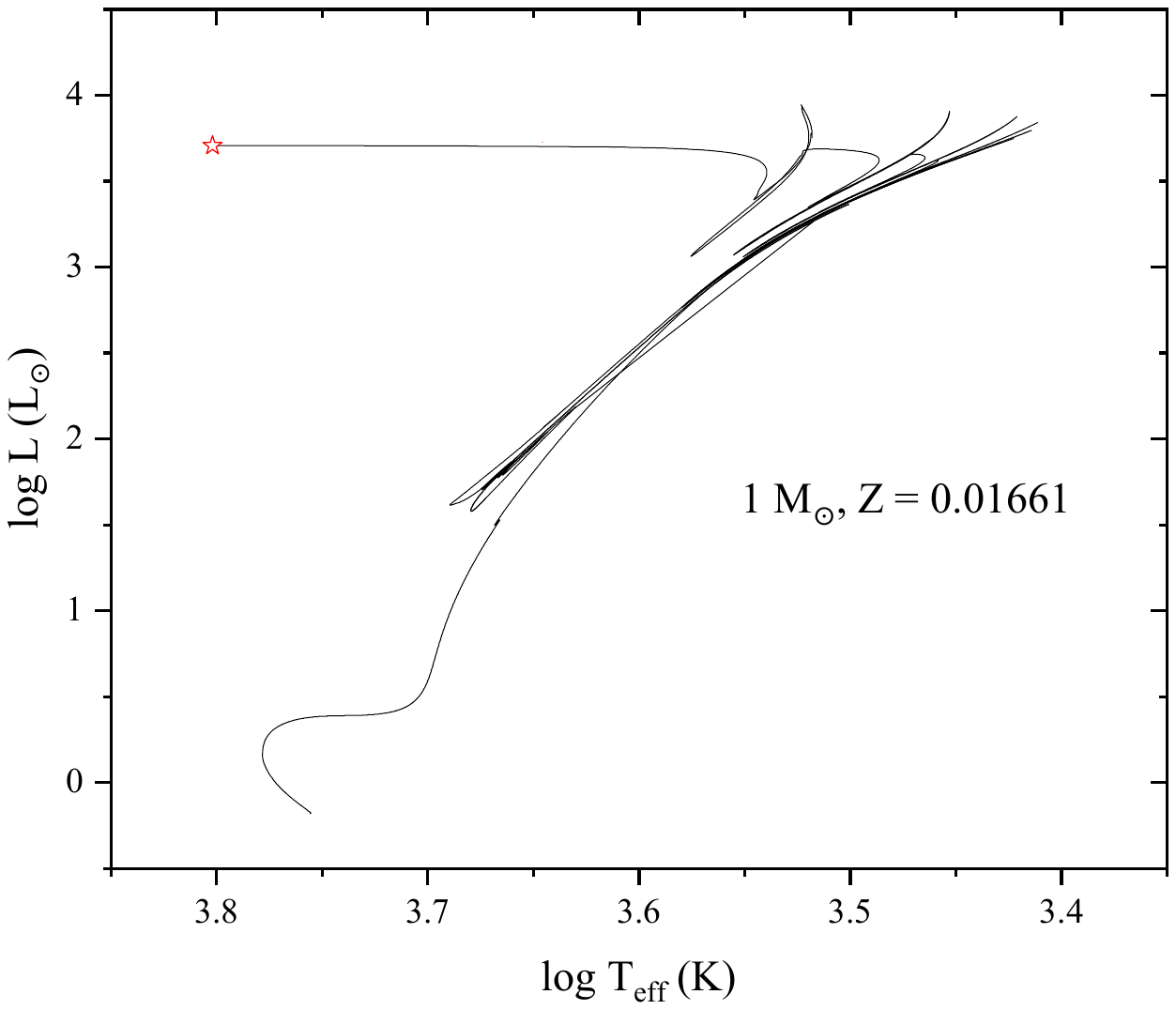}
\caption{ 
Evolution in the HRD of the $1 \> \mathrm{M}_\odot$ model from the ZAMS up to the onset of the EPAGBI (open star).}
\label{fig:figure1}
\end{figure}

In figure \ref{fig:figure1}, we show the evolutionary track in the Hertzsprung-Russell diagram (HRD) of our $1 \> \mathrm{M}_\odot$ model from the Zero Age Main Sequence (ZAMS) up to the point at which the EPAGBI begins (marked by the open star). This model experiences 9 helium flashes on the thermally pulsing asymptotic giant branch. Some properties of the model on the ZAMS are given in the first line of  table \ref{tab:table1}. Here, $M_H$ and $M_{He}$ are the total hydrogen mass and total helium mass, respectively. The photospheric mass fractions of H, He, $^{12}\mathrm{C}$, $^{14}\mathrm{N}$, and $^{16}\mathrm{O}$, are $X_\mathrm{H}, X_{\mathrm{He}}, X_{^{12}\mathrm{C}}, X_{^{14}\mathrm{N}}$ and $X_{^{16}\mathrm{O}}$, respectively. The final column shows the photospheric lithium abundance, $X_{^{7}\mathrm{Li}}$. The same properties of the model after it has left the AGB and just before the onset of the EPAGBI are given in the second line of the table. At this stage, the surface lithium abundance corresponds to depletion by a factor of 36 relative to the initial pre-main sequence lithium abundance.

\begingroup
\renewcommand{\arraystretch}{1.5} 
\begin{table*}
\centering
\caption{Properties of $1 \> \mathrm{M}_\odot$ models at specific evolutionary stages}
\label{tab:table1}
\setlength{\tabcolsep}{4pt}
\begin{tabular}{l|cccccccccc}  
\hline
Stage  &  $M$  & $M_\mathrm{H}$ & $M_{\mathrm{He}}$ & $X_\mathrm{H}$ &  $X_{\mathrm{He}}$ & $X_{^{12}\mathrm{C}}$ & $X_{^{14}\mathrm{N}}$ & $X_{^{16}\mathrm{O}}$ & $X_{^7\mathrm{Li}}$ \\ 
             & ($\mathrm{M}_\odot$)     & ($\mathrm{M}_\odot$)  & ($\mathrm{M}_\odot$)  &        &      &  	 &   	 \\ \hline
ZAMS                & 1.000 &  0.713                 & 0.270                 &  0.713 & 0.270 & 2.84 $\times 10^{-3}$ & 8.83 $\times 10^{-4}$ & 8.01 $\times 10^{-3}$ 
 & 9.96 $\times 10^{-9}$  \\
AGB departure       & 0.567 &  6.11 $\times 10^{-4}$ & 1.55 $\times 10^{-2}$ &  0.683 & 0.298 & 3.17 $\times 10^{-3}$ & 1.55 $\times 10^{-3}$ & 7.96 $\times 10^{-3}$ 
 & 2.79 $\times 10^{-10}$ \\
 Hydrodynamic       & 0.567 &  2.91 $\times 10^{-4}$ & 1.22 $\times 10^{-2}$ &  0.400 & 0.545 & 1.53 $\times 10^{-2}$ & 2.05 $\times 10^{-2}$ & 1.17 $\times 10^{-2}$ & 2.18 $\times 10^{-8}$ \\
 CSPN               & 0.566  & 1.31 $\times 10^{-4}$ & 1.13 $\times 10^{-2}$ &  0.053 & 0.342 & 0.383                 & 1.10 $\times 10^{-2}$ & 0.0178 & 2.71 $\times 10^{-9}$ \\
 WD start           & 0.565  & 5.52 $\times 10^{-6}$ & 6.32 $\times 10^{-3}$ &  0.045 & 0.326 & 0.400                 & 9.71 $\times 10^{-3}$ & 0.0186 & 1.99 $\times 10^{-16}$ \\
\hline
\end{tabular}
\centering						       
\end{table*}
\endgroup

\begin{figure}
\centering
\includegraphics[scale=0.3]{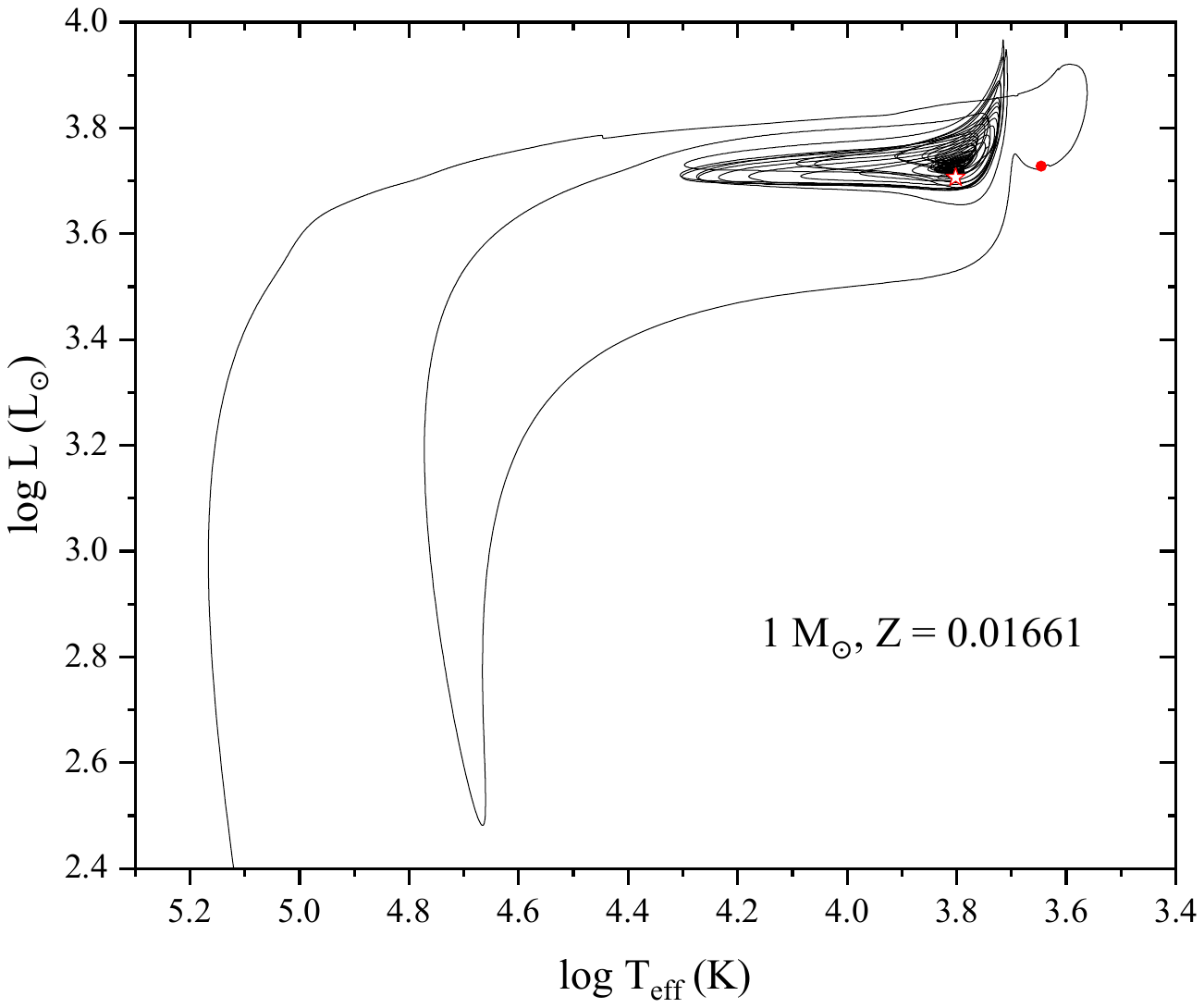}
\caption{ 
Evolution in the HRD of the $1 \> \mathrm{M}_\odot$ model from the onset of the EPAGBI (open star) to the early part of the white dwarf cooling track. The red circle shows the point at which the envelope becomes hydrodynamically unstable.}
\label{fig:figure2}
\end{figure}

\begin{figure}
\centering
\includegraphics[scale=0.3]{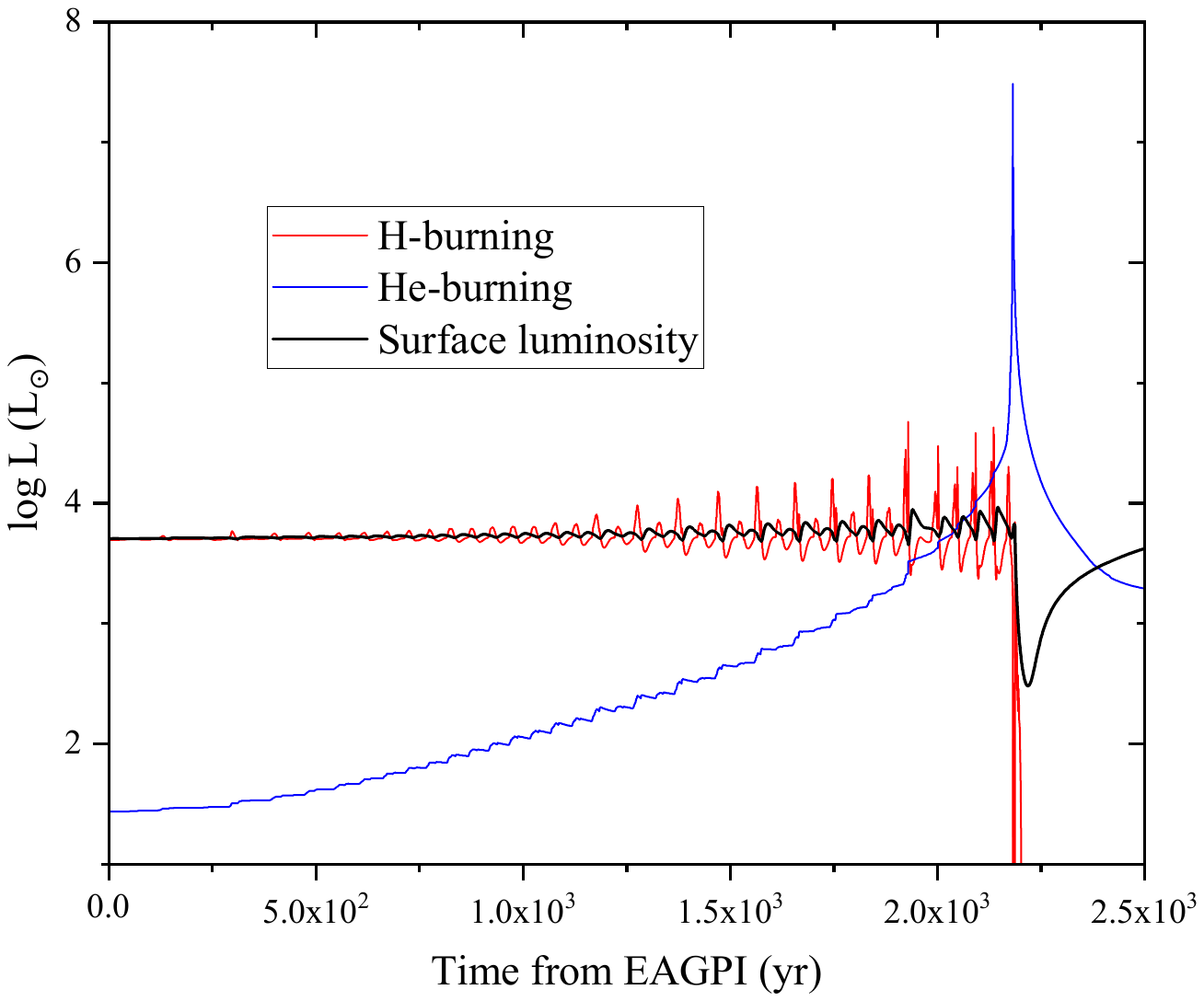}
\caption{ 
Evolution of the hydrogen and helium burning luminosities together with the stellar luminosity from the onset of the EPAGBI in the $1 \> \mathrm{M}_\odot$ model.}
\label{fig:figure3}
\end{figure}

 In figure \ref{fig:figure2}, we show the evolution in the HRD from the onset of the EPAGBI (marked by the open star) to the early stages of the WD cooling track. The loops are due to thermal instability of the hydrogen burning shell. The hydrogen and helium burning luminosities are shown together with the stellar luminosity in figure \ref{fig:figure3}. Initially the instability is oscillatory with a period of about 5.5 yr with amplitude that exponentially increases on a time scale of 400 yr. After 1000 yr from onset there is a period doubling, with a transition to slightly chaotic behavior at about 1900 yr. During the phase in which the hydrogen burning pulses, the helium burning luminosity increases leading to a thermal pulse which peaks at about 2200 yr. This pulse leads to the large loop in figure \ref{fig:figure2}, which is similar in behavior to an AGB final thermal pulse (AFTP) \citep{2001Ap&SS.275....1B}. When the model returns to the red, it becomes unstable (at the point shown by the red circle in figure \ref{fig:figure2}). The instability is due to the luminosity exceeding the local Eddington luminosity in a region near the stellar surface  where the density is too low for convection to efficiently transport energy. As shown by \cite{1973ApJ...181..429J}, this leads to a density inversion. Outside the density inversion, the opacity is low and radiation pressure is negligible. This leads to a shell of material supported by gas pressure. With further evolution, the mass in the shell increases up to a point where it cannot be supported by gas pressure and it rapidly collapses inwards in a hydrodynamic matter. We find that the evolution can be continued by forcing the outer parts of the star, i.e. where the temperature is less than $10^6$ K, to be in thermal equilibrium. The final white dwarf mass is $0.565 \> \mathrm{M}_\odot$ in excellent agreement with the semi-empirical 'MIST-based' initial - final mass relation of \cite{2018ApJ...866...21C}, which gives $M_f = 0.569 \pm 0.046 \> \mathrm{M}_\odot$.

During the HRD loops, the model oscillates between spectral types A and B. The surface temperature of BA stars are such that they can have two or three convection zones (due to opacity peaks associated with H, He$^+$ and He$^{++}$ ionization zones and the iron opacity bump at $T = 200,000$ K). This leads to an intricate convection zone structure that results in dredge-up of heavy elements.

\begin{figure}
\centering
\includegraphics[scale=0.3]{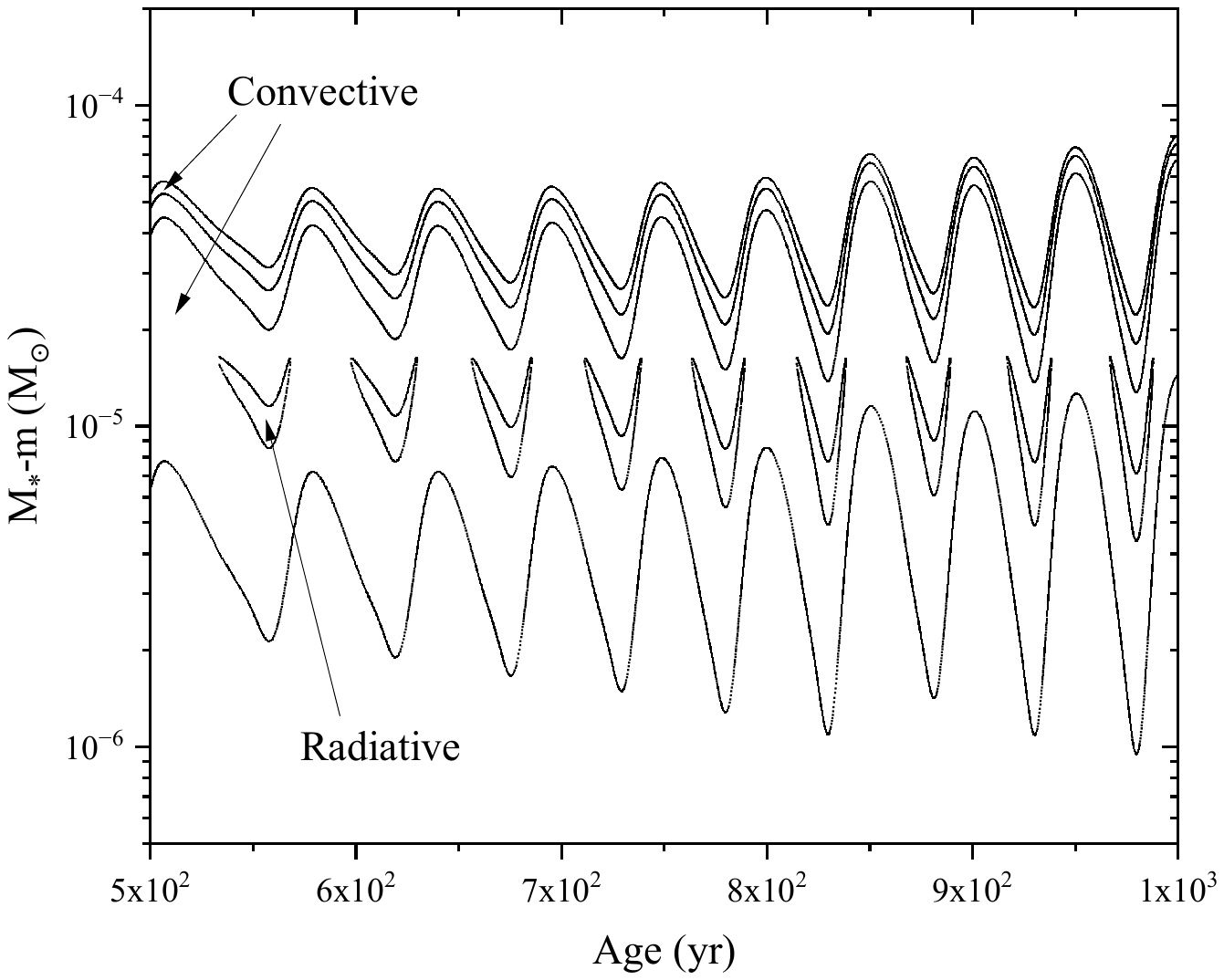}
\caption{ 
Convection zone evolution from 500 to 1000 yr after the onset of the EPAGBI in the $1 \> \mathrm{M}_\odot$ model.}
\label{fig:figure4}
\end{figure}

 In figure \ref{fig:figure4}, we show the location of convection zone boundaries in terms of mass measured from the stellar surface as a function of time from the onset of the EPAGBI. For figure \ref{fig:figure4}, the time span is from 500 to 1000 yr after the onset of the EPAGBI. During this time interval, the number of convection zones alternates between 2 and 3. 

\begin{figure}
\centering
\includegraphics[scale=0.3]{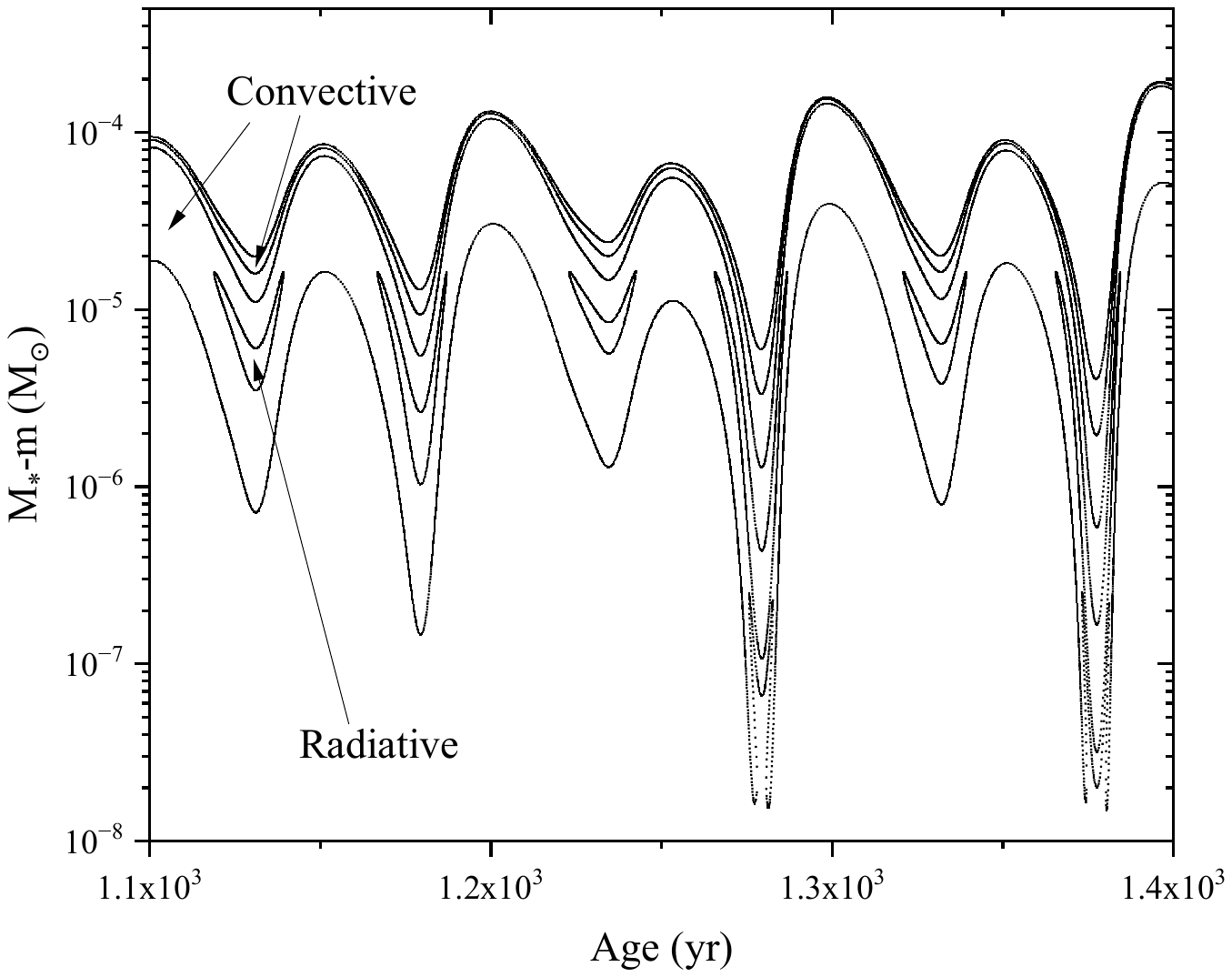}
\caption{ 
Convection zone evolution from 1100 to 1400 yr after the onset of the EPAGBI in the $1 \> \mathrm{M}_\odot$ model.}
\label{fig:figure5}
\end{figure}

\begin{figure}
\centering
\includegraphics[scale=0.3]{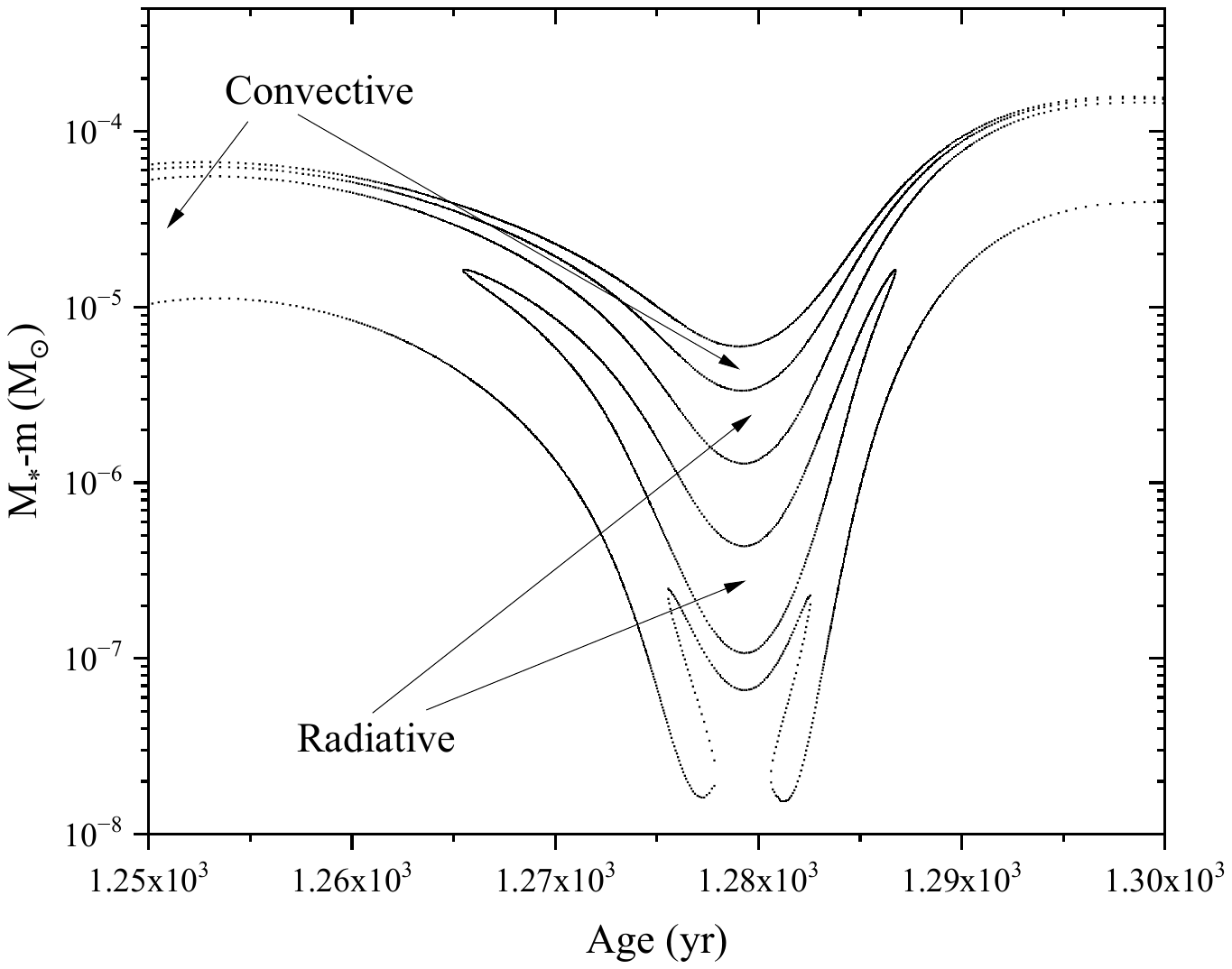}
\caption{Convection zone evolution from 1250 to 1300 yr after the onset of the EPAGBI in the $1 \> \mathrm{M}_\odot$ model.}
\label{fig:figure6}
\end{figure}

 Figure \ref{fig:figure5} is similar to figure \ref{fig:figure4} except for a time interval from 1100 to 1400 yr (prior to which the period doubling has occurred). During this interval there can be as many as 4 convection zones. This can be seen more clearly in figure \ref{fig:figure6} which covers the interval 1250 to 1300 yr.
 
 In typical thermal pulse fashion, shortly after the helium-flash driven convection ends, the base of the surface convection zone moves inwards to a region enriched in elements synthesized during helium burning. Because $M_H$ at this point is $\sim 10^{-3} \> \mathrm{M}_\odot$ and the total mass in the convection zone is $\sim 10^{-2} \> \mathrm{M}_\odot$, the hydrogen mass fraction in the convection zone becomes significantly reduced. The third row of Table 1 gives the stellar properties at the point when hydrodynamic behavior ensues.  The increase in heavy element abundances leads to large  mass loss rates due to radiatively driven winds. Envelope properties at the beginning of the CSPN phase [which, following \cite{1984ApJ...277..333I}, we take to be when $T_\mathrm{eff}$ = 30,000 K] are given on line 4 of table 1, The photospheric mass fractions are similar to those of [WCL] CSPNe \citep{2024arXiv241118035M}. Even if the envelope is not ejected hydrodynamically, further mass loss reduces $M_H$ to $ \sim 5 \times 10^{-6} \> \mathrm{M}_\odot$ by the time the model reaches $T_\mathrm{eff} = 150,000$ K at the top of the WD cooling track (line 5 of Table 1). Due to the convective mixing episode that follows the AFTP, the hydrogen is thoroughly mixed in a layer of heavier elements of mass $10^{-4} \> \mathrm{M}_\odot$.

 At the point when hydrodynamic behavior ensues, the lithium abundance has been enhanced by a factor of 2.2 relative to the initial lithium abundance, due to operation of the Cameron-Fowler mechanism \citep{1971ApJ...164..111C}. During the later HRD loops, the base of the convection zone reaches temperatures greater than $10^7$ K where significant amounts of $^7$Be are produced by the $^3$He($\alpha$,$\gamma$)$^7$Be reaction, some of which is convectively transported outward to cool enough temperatures  ($ < 3 \times 10^6$ K) where $^7$Li produced by the $^7$Be(,e$^-$$\nu$)$^7$Li reaction can survive.
 
\subsection{Evolution of the \texorpdfstring{$1 \> \mathrm{M}_\odot$}{} model with EPAGBI suppressed}

In previous studies modeling the evolution of low-mass stars toward the WD stage, dynamical instabilities on the AGB have been occasionally reported but there has been no mention of the EPAGBI until the study by \cite{2023arXiv230311374G}. There are a number of possible reasons. Early work e.g. by Iben and collaborators used a code that used separate thermal equilibrium envelope calculations \citep{1965ApJ...141..993I}. The instability can also be suppressed if the time step is chosen to be sufficiently large (e.g. longer than the pulsation period) even when thermal equilibrium in the outer layers is not assumed.

For purposes of comparison with the results of the previous section, we have followed the evolution when the EAGBI is suppressed by forcing the time step to be greater than 10 years from the point at which $T_\mathrm{eff} = 6,500$ K to the point where radiative driven winds begin to dominate the mass loss  at $T_\mathrm{eff} = 9,700$ K. The minimum allowed time step is then reduced to 1 yr. In comparison, the time step during the EPAGBI phase is typically 0.01 to 0.1 yr. The time step exceeds the enforced minimum when $T_\mathrm{eff} = 14,300$ K and so we then remove the constraint on the time step. The model enters the WD cooling phase with a hydrogen-rich atmosphere. When the surface luminosity $\sim 750 \> \mathrm{L}_\odot$, a VLTP occurs, during which the hydrogen burning luminosity reaches $4.1 \times 10^{10} \> \mathrm{L}_\odot$. The high energy generation rate causes the model to rapidly expand, doubling in size in a few minutes. The evolution becomes hydrodynamic due to the velocity in the surface layers exceeding the local sound speed, and our hydrostatic code can no longer follow the evolution. As discussed previously, to allow evolution beyond the VLTP, we found it necessary to force the outer layers ($T < 10^6$ K) to be in thermal equilibrium.  At the beginning of the second descent of the white dwarf cooling track, the total mass of remaining hydrogen is $6.3 \times 10^{-7} \> \mathrm{M}_\odot$ uniformly distributed in a layer of mass $\sim 2.0 \times 10^{-5} \> \mathrm{M}_\odot$ of abundances $X_\mathrm{H} = 6.16 \times 10^{-3}, X_\mathrm{He} = 0.444, X_{^{12}\mathrm{C}} = 0.346, X_{^{13}\mathrm{C}} = 4.47 \times 10^{-2}, X_{^{14}\mathrm{N}} = 1.74 \times 10^{-2}$ and $X_{^{16}\mathrm{O}} = 0.100$. Further mass loss reduces the hydrogen mass to $7.39 \times 10^{-8} \> \mathrm{M}_\odot$.

\begin{figure}
\centering
\includegraphics[scale=0.3]{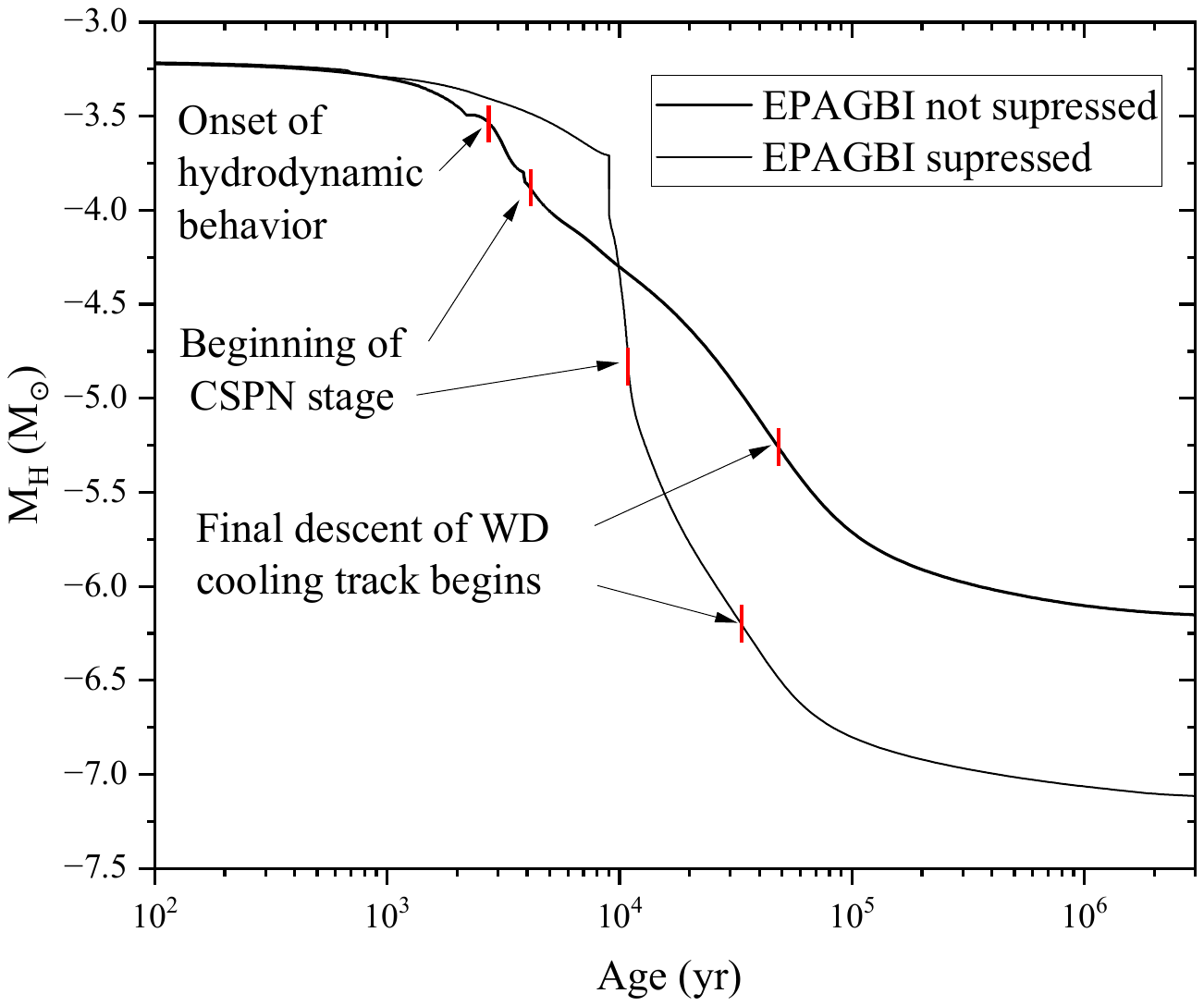}
\caption{Evolution of $M_H$ from the onset of the EPAGBI for the $1 \> \mathrm{M}_\odot$ models.}
\label{fig:figure7}
\end{figure}

Figure \ref{fig:figure7} shows how $M_H$ evolves with the time from the onset of the EPAGBI for the two cases with and without suppression of the EPAGBI. The red vertical marks correspond to evolutionary stages given in table \ref{tab:table2}. We see that the EPAGBI does not lead to significant additional mass loss (compared to calculations in which the the EPAGBI is suppressed), and hence does not play much of a role in determining the final hydrogen layer mass.

In both simulations, with and without suppression of the EPAGBI, the mass of hydrogen at the beginning of the WD cooling phase is significantly less than the canonical value of $10^{-4} \> \mathrm{M}_\odot$ for a DA WD, and is consistent with hydrogen-layer masses determined from asteroseismological analysis. However, hydrodynamic behavior is found to occur in the later stages of both simulations when there is still strong helium burning, and it is possible that most of the hydrogen layer is dynamically ejected. Hence the masses predicted from hydrostatic models should be considered as upper estimates.

\subsection{Evolution of the \texorpdfstring{$2 \> \mathrm{M}_\odot$}{} model}

\begin{figure}
\centering
\includegraphics[scale=0.3]{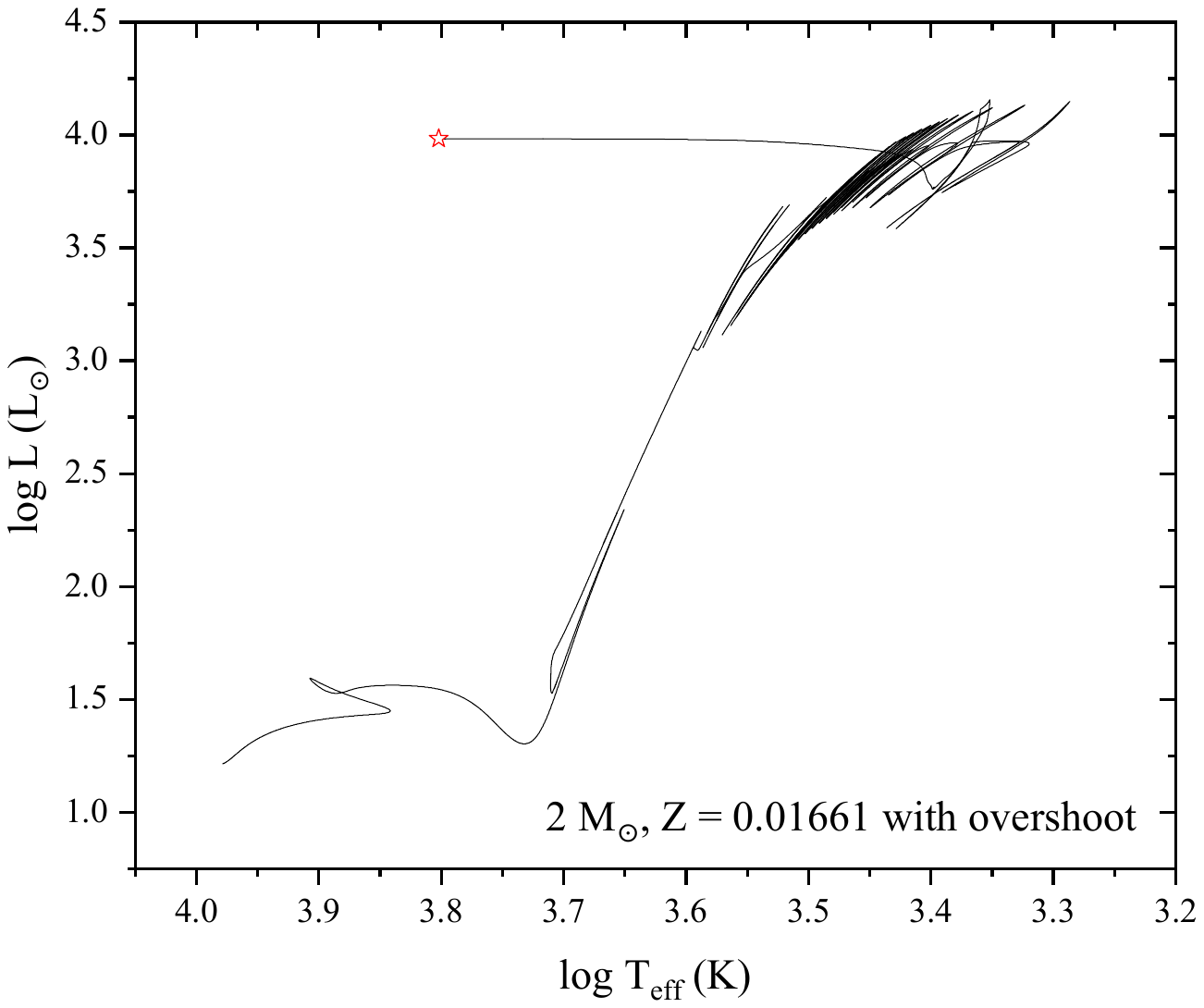}
\caption{ 
Evolution in the HRD of the $2 \> \mathrm{M}_\odot$ model from the ZAMS up to the onset of the EPAGBI (open star).}
\label{fig:figure8}
\end{figure}

In figure \ref{fig:figure8}, we show the evolutionary track in the Hertzsprung-Russell (HRD) of our $2 \> \mathrm{M}_\odot$ model up to the point at which the EPAGBI begins. In this model, convective overshoot was included. We found that if convective overshoot was not included but all other parameters were kept the same, thermal instability of the envelope occurred during the late stages of the Thermally Pulsating AGB (TPAGB). The model with convective overshoot experiences a dual shell flash \citep{2008A&A...490..769C} followed by 25 helium flashes on the thermally pulsing asymptotic giant branch. The envelope properties at the ZAMS and the beginning of the EPAGB are given in lines 1 and 2 of table \ref{tab:table2}, respectively.  The lithium abundance corresponds to enhancement by a factor of 8 relative to the initial lithium abundance.

\begingroup
\renewcommand{\arraystretch}{1.5} 
\begin{table*}
\centering
\caption{Properties of $2 \> \mathrm{M}_\odot$ models at specific evolutionary stages}
\label{tab:table2}
\setlength{\tabcolsep}{4pt}
\begin{tabular}{l|cccccccccc}  
\hline
Stage  &  $M$  & $M_\mathrm{H}$ & $M_{\mathrm{He}}$ & $X_\mathrm{H}$ &  $X_{\mathrm{He}}$ & $X_{^{12}\mathrm{C}}$ & $X_{^{14}\mathrm{N}}$ & $X_{^{16}\mathrm{O}}$ & $X_{^7\mathrm{Li}}$ \\  
             & ($\mathrm{M}_\odot$)     & ($\mathrm{M}_\odot$)  & ($\mathrm{M}_\odot$)  &        &      &  	 &   	 \\ \hline
ZAMS                & 2.000 &  1.426                 & 0.541                 &  0.713 & 0.270 & 2.84 $\times 10^{-3}$ &  8.83 $\times 10^{-4}$ & 8.01 $\times 10^{-3}$ & 9.96 $\times 10^{-9}$ \\
AGB departure       & 0.642 &  1.99 $\times 10^{-4}$ & 1.03 $\times 10^{-2}$ &  0.646 & 0.323 & 1.20 $\times 10^{-2}$ & 2.83 $\times 10^{-3}$ & 8.45 $\times 10^{-3}$ 
 & 8.09 $\times 10^{-8}$ \\
 End                & 0.640 & 3.51 $\times 10^{-7}$ & 6.96 $\times 10^{-3}$ &  1.2 $\times 10^{-4}$ & 0.336 & 0.402                 & 4.44 $\times 10^{-3}$ & 0.176 & 4.28 $\times 10^{-7}$ \\
 End - EPAGBI suppressed & 0.641 & 9.15 $\times 10^{-7}$ & 8.50 $\times 10^{-3}$ &  4.4 $\times 10^{-4}$ & 0.518 & 0.254            & 3.39 $\times 10^{-2}$ & 9.05 $\times 10^{-2}$ & 3.41 $\times 10^{-7}$ \\
\hline
\end{tabular}
\centering						       
\end{table*}
\endgroup

\begin{figure}
\centering
\includegraphics[scale=0.3]{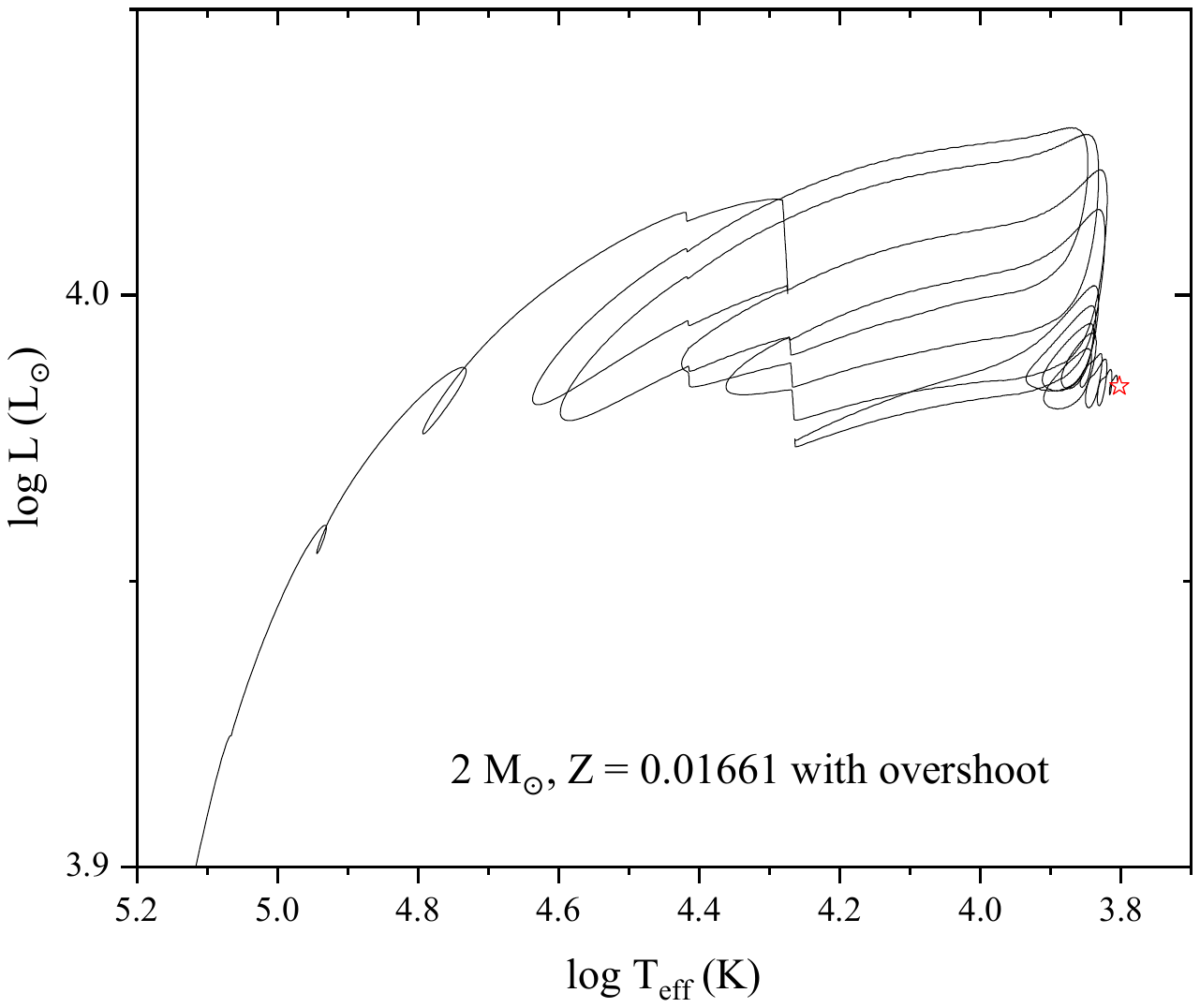}
\caption{ 
Evolution in the HRD of the $2 \> \mathrm{M}_\odot$ model from the onset of the EPAGBI (open star) to just before the white dwarf cooling track.}
\label{fig:figure9}
\end{figure}

 Figure \ref{fig:figure9} shows the evolution in the HRD from the start of the EPAGBI phase to near the beginning of the WD cooling track. The loops  due to thermal instability of the hydrogen burning shell are clearly visible. The near vertical parts of the track near log $T_\mathrm{eff}$ = 4.3 and 4.4 are due to the bi-stability jumps in the wind mass loss rate, that result from increased Fe III line acceleration interior to the sonic point  \citep{1999A&A...350..181V}.

\begin{figure}
\centering
\includegraphics[scale=0.3]{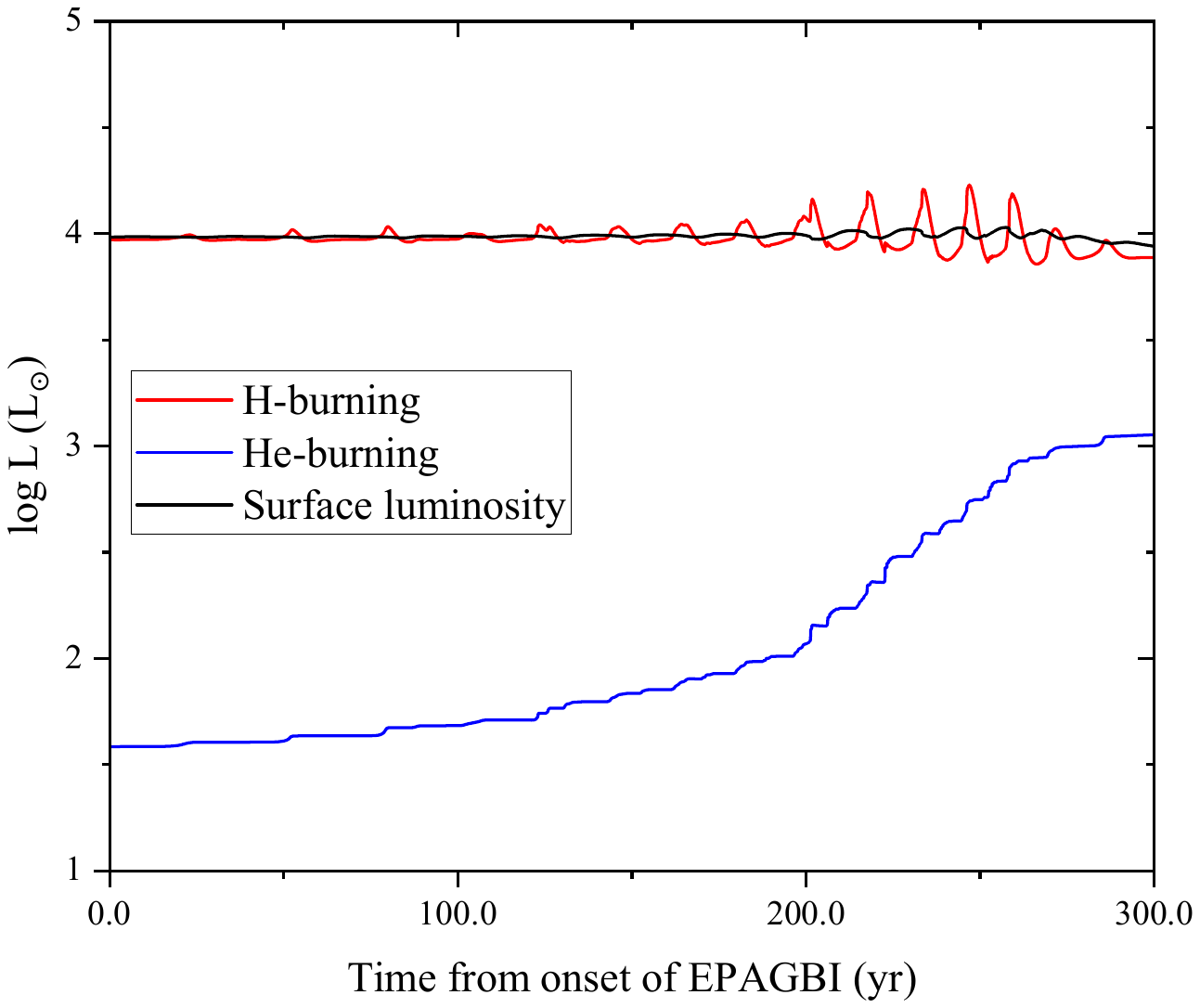}
\caption{ 
Evolution of the hydrogen and helium burning luminosities together with the stellar luminosity from the onset of the EPAGBI in the $2 \> \mathrm{M}_\odot$ model.}
\label{fig:figure10}
\end{figure}

The hydrogen and helium burning luminosities are shown together with the stellar luminosity in figure \ref{fig:figure10}. The oscillation period is initially about 30 yr shortening to about 15 yr as the amplitude increases on a time scale of $\sim 400$ yr, before the oscillation dies away. During the phase in which the hydrogen burning pulses, the helium burning luminosity does increases but, unlike the evolution of the $1 \> \mathrm{M}_\odot$ model, not yet to the level of a thermal pulse.

\begin{figure}
\centering
\includegraphics[scale=0.3]{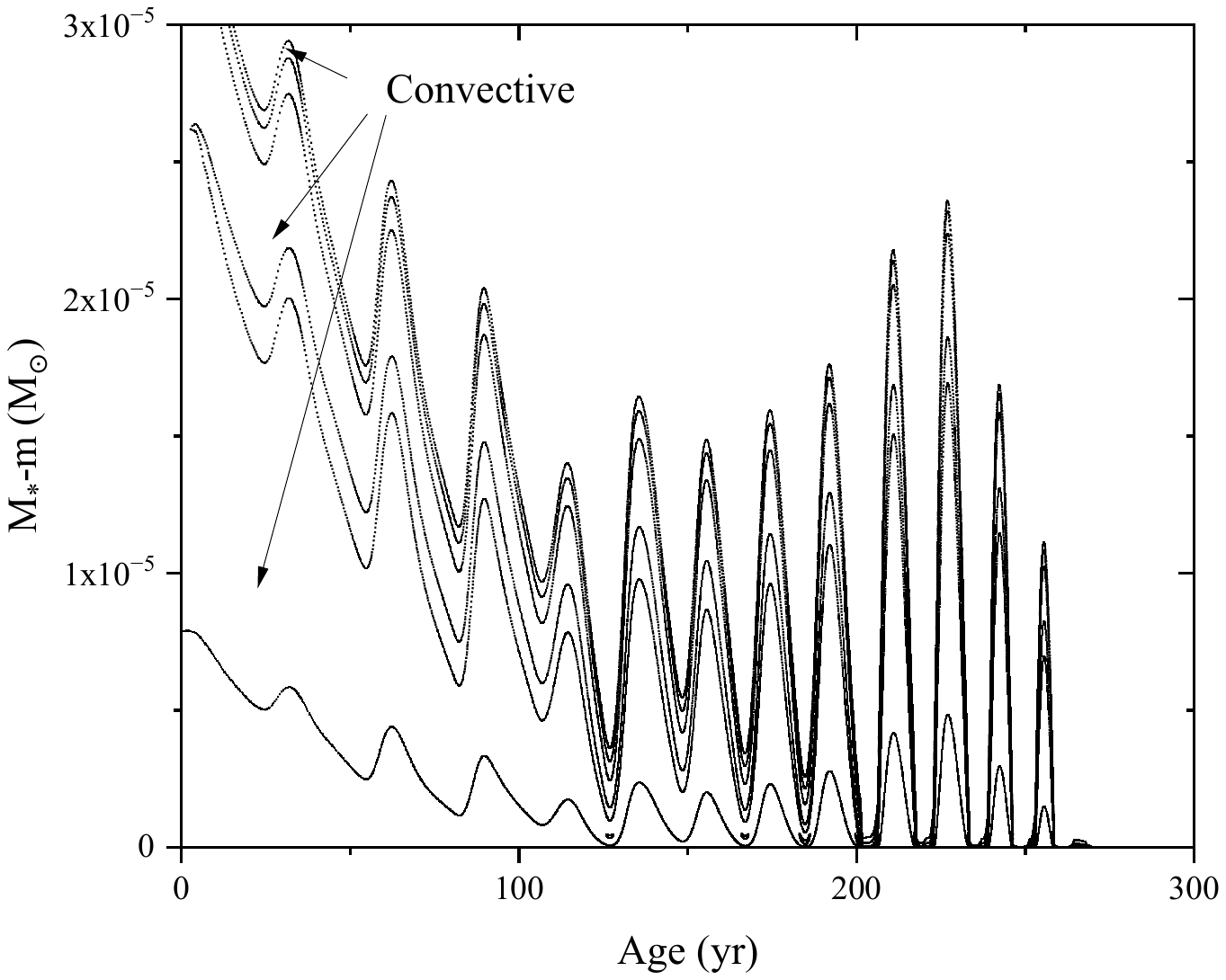}
\caption{ 
Convection zone evolution from 0 to 250 yr after the onset of the EPAGBI in the $2 \> \mathrm{M}_\odot$ model.}
\label{fig:figure11}
\end{figure}

In figure \ref{fig:figure11}, we show the convection zones as a function of time from the onset of the EPAGBI. Unlike the $1 \> \mathrm{M}_\odot$ case, the number of convection zones remains constant with a value of 3 until the surface becomes too hot for hydrogen recombination to occur.

\begin{figure}
\centering
\includegraphics[scale=0.3]{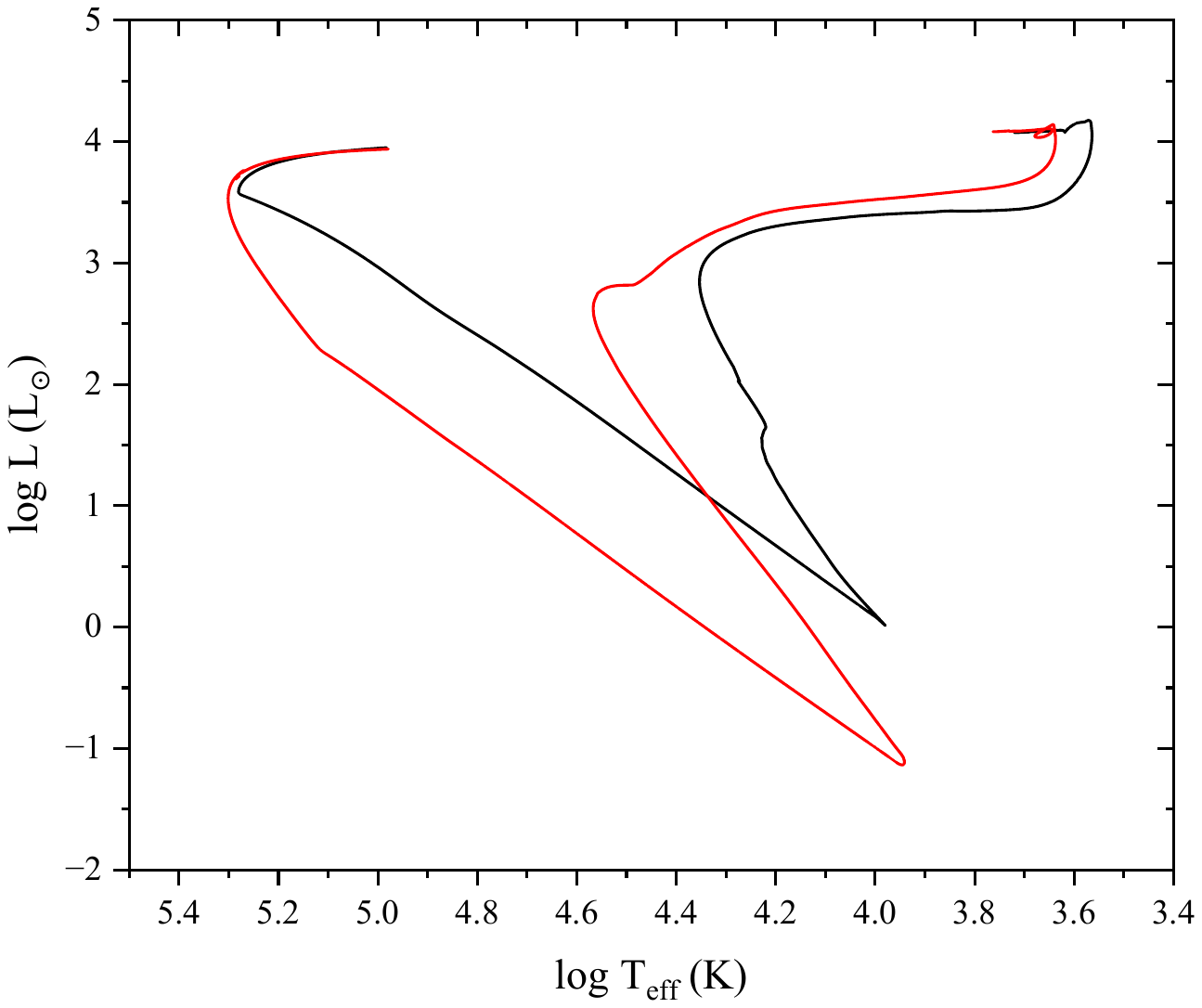}
\caption{ 
Post EPAGBI evolution of the $2 \> \mathrm{M}_\odot$ models in the HRD. The black and red lines are for the cases in which the EPAGBI is included and suppressed, respectively}
\label{fig:figure12}
\end{figure}

From figure \ref{fig:figure10}, it can be seen that the helium burning luminosity increases during the phase of instability. The helium burning luminosity continues to increase and a VLTP occurs. After a time 680 yr from the start of the EPAGBI, the helium burning luminosity reaches a maximum of $6 \times 10^7 \> \mathrm{L}_\odot$. At this point the convection zone driven by the helium shell flash reaches the hydrogen rich layer. Initially $L_{He}$ decreases but convective mixing of protons into the helium layer later leads to increases in both $L_{He}$ and $L_{H}$, which reach maxima of $8 \times 10^8 \> \mathrm{L}_\odot$ and $L_{H}$ to $9 \times 10^8 \> \mathrm{L}_\odot$, respectively. The star expands back to giant dimensions on a timescale of order 10 yr. During the DSF, heavy elements are dredged to the surface, and reach the photosphere during the return to the giant branch. When the model evolves to the blue for the second time, it does so with a higher luminosity than the first time and also with an atmosphere that has been significantly enriched in heavy elements ($Z = 0.48$). The higher luminosity and opacity leads to the Eddington luminosity being exceeded in the envelope, which causes a density inversion where the gradient in gas pressure is inward in opposition to the radiation pressure gradient \citep{1973ApJ...181..429J, 1998A&A...330..641A, 2003ApJ...583..913L}. Once the density becomes too low for efficient convective energy transport to occur, the outer layers become unstable and the evolution can no longer be followed with our hydrostatic equilibrium code. A further complication is that the low-temperature opacity tables do not cover high enough Z values. The evolution of our 2 $\> \mathrm{M}_\odot$ model in the HRD is shown by the black line in figure \ref{fig:figure12}.

The envelope properties at the end of the calculation, are given in line 3 of table \ref{tab:table2}. The lithium abundance corresponds to enhancement by a factor of 42 relative to the initial lithium abundance. Significant amounts of $^{13}$C are produced during the VLTP and are convectively transported to the surface. At the end of the calculation, the surface $^{13}$C mass fraction is $2.12 \times 10^{-2}$. The production of neutrons by the $^{13}$C($\alpha$,n)$^{16}$O reaction is expected to result in formation of s-process nuclei.

At the end of the calculation, the hydrogen, of total mass $M_H$ = $3.51 \times 10^{-7} \> \mathrm{M}_\odot$,  is distributed in a helium and carbon-rich layer of mass $2.8 \times 10^{-3} \> \mathrm{M}_\odot$.

\subsection{Evolution of the \texorpdfstring{$2 \> \mathrm{M}_\odot$}{} model with EPAGBI suppressed}
When the EPAGBI is suppressed, the model also experiences a VLTP but it occurs further down the cooling track than when the EPAGBI is not suppressed. This leads to a stronger hydrogen flash with maximum values of $L_{He}$= $2.0 \times 10^8 \> \mathrm{L}_\odot$ and $L_{H}$ = $3.4 \times 10^{10} \> \mathrm{L}_\odot$. The evolutionary path in the HRD for this model is shown by the red line in figure \ref{fig:figure12}.

To allow evolution beyond the VLTP, we again found it necessary to force the outer layers ($T < 10^6$ K) to be in thermal equilibrium. 
Again, when the model evolves to the blue for the second time, a density inversion occurs, leading to instability of the outer layers.
The envelope properties at the end of the calculation are given in line 4 of table \ref{tab:table2}. The lithium abundance corresponds to enhancement by a factor of 34 relative to the initial lithium abundance. The hydrogen, of total mass $M_H = 9.16 \times  10^{-7} \> \mathrm{M}_\odot$,  is distributed in a helium and carbon-rich layer of mass  $2.8 \times 10^{-3} \> \mathrm{M}_\odot$.

\section{Discussion and Conclusions}

We have presented results for the post-AGB evolution of solar composition models of mass $1$ and $2 \> \mathrm{M}_\odot$. As first found by \cite{2023arXiv230311374G}, rapidly growing radial pulsations develop due to thermal instability of the hydrogen burning shell. The pulsations manifest as loops in the HRD. For comparison purposes, we have calculated the evolution of the same models but with the instability suppressed by using time steps comparable to the thermal timescale.

In both evolutionary calculations, with and without suppression of the EPAGBI, we find that VLTPs occur. This is probably due to a selection effect in that we only consider models that can be followed to the post-AGB phase. Other models experience instability at an earlier stage in the evolution of the TPAGB. \cite{1994A&A...290..807W} found that models with mass in the range $\simeq 1 - 4 \> \mathrm{M}_\odot$ experienced a runaway expansion during the luminosity peak of a thermal pulse once the models exceed a critical luminosity, and attribute the expansions to recombination of hydrogen within a thermally and dynamically unstable envelope. 

Because of the VLTPs, mass loss reduces the mass of hydrogen to much lower values than the canonical stellar evolution value of $10^{-4} \> \mathrm{M}_\odot$ for a DA WD star. The  $M_H$ values at the end of the calculations are in the range consistent with asteroseismological determinations, but we caution that, because hydrodynamic behavior is not included, it is possible that all hydrogen would be removed. Thus, it is unclear whether the occurrence of the EPAGBI resolves the discrepancy between predictions of stellar evolution modeling and the asteroseismological hydrogen envelope mass determinations. Any additional mass loss during the EPAGBI stage is insignificant when compared to the total mass loss that occurs the end of the AGB. The major impact of EPAGBIs is that they cause loops in the HRD that are absent when they are suppressed. For models of AGB-departure mass 0.567 and 0.642 $\mathrm{M}_\odot$, it takes approximately 100 and 10 yr for a single HRD loop, respectively. Such loops might be detectable in a long-term monitoring program, or perhaps by their imprint on planetary nebula morphology imparted by the cyclically varying mass loss rate. Since the characteristic timescale of the looping in the HRD depends on the stellar mass, if measurable, it could provide a way to determine the stellar mass just after AGB departure, particularly if it is $\gtrsim 0.72 \> \mathrm{M}_\odot$ for which we estimate a 1 yr loop timescale. 

Another signature of the EPAGBI is the production of lithium by the Cameron-Fowler process. During the EPAGBI phase the photospheric temperature  is always greater than 6,300 K, which is much higher than the temperatures of stars of appropriate log \textit{g} for which the Li I 670.8 nm resonance line can be detected, [e.g. \citep{2018MNRAS.478.4513B}]. Hence, Li detection is unlikely to be a way to identify the EPAGBI phase. However, the Li precursor, $^7$Be, is convected to the photosphere in significant amounts (up to $X_{^7\mathrm{Li}}$ $\sim 3 \times 10^{-8}$, which is $\sim 400$ times the solar photospheric mass fraction) at various times in the EPAGBI phase, and may be detectable by observing the Be II 313.0 and 313.1 nm resonance doublet, which has been identified in the Sun and stars of F and G spectral type [e.g. \citep{2022ApJ...941...21B}].

Instabilities similar to the EPAGBI might also be expected to occur when a low-mass star leaves the RGB, either before or shortly after the helium-core flash, due to mass loss at rates higher than the typical Reimers mass loss rate, $\eta_r = 0.477$ \citep{2015MNRAS.448..502M}. \cite{2023MNRAS.525.4700L} found such behavior in models of Population III stars for masses $\lesssim 0.9 \> \mathrm{M}_\odot$ which experience significant convective dredge-up of $^{12}$C produced in a double core flash \citep{2008A&A...490..769C}. The consequent increase in envelope opacities leads to expansion of the star and increased mass loss rates due to a dusty wind. Instability is also expected to occur in the ‘hot flasher’ scenario  \citep{1993ApJ...407..649C}, used by \cite{1996ApJ...466..359D} to create extreme horizontal branch (EHB) stars by increasing the standard Reimers' mass loss rate by a factor of order 2. Indeed, we find strong instabilities in the evolution of models of initial mass 0.85 $\mathrm{M}_\odot$ and heavy element abundance Z = $2 \times 10^{-3}$ when $\eta_r$ is set to 0.91. However, \cite{2023MNRAS.525.4700L} did not find instability in their Pop. III models. We have recalculated the post-AGB evolution of their 0.85 $\mathrm{M}_\odot$ model with a shorter time step, and found evidence for weak instability that does not grow, and certainly does not lead to  HRD loops. The major difference in Pop. III models from higher metallicity models is that there is no iron-opacity bump. \cite{2023arXiv230311374G} found that the pulsation driving is confined to the hydrogen and helium partial ionization zones, and at the location of the bottom of the Z opacity bump, the pulsation is strongly damped. One might then expect the instability to be more strongly excited in Pop. III models, the opposite of what was found by \cite{2023MNRAS.525.4700L}. \cite{2023arXiv230311374G} also noted that existence of the instability  might depend on whether the assumption of instantaneous convective adjustment is valid, as the convection timescales in the driving regions are comparable to the oscillation timescales. 

\section*{Acknowledgments}

The author thanks Conor Larsen and the anonymous referee for providing valuable comments on the manuscript, and acknowledges stimulating conversations with Judi Provencal and Henry Sanford-Crane.

\bibliography{refs}
\bibliographystyle{aasjournal}

\end{document}